\documentclass{revtex4}
\usepackage{amsmath}
\usepackage{graphicx}
\newcommand{\be}{\begin{equation}}
\newcommand{\ee}{\end{equation}}
\begin{document}
\title{Random Unitaries Give Quantum Expanders}
\author{M.~B.~Hastings}
\affiliation{Center for Nonlinear Studies and Theoretical Division,
Los Alamos National Laboratory, Los Alamos, NM, 87545}
\begin{abstract}
We show that randomly choosing the matrices in a completely positive map from
the unitary group gives a quantum expander.  We consider Hermitian and
non-Hermitian cases, and we provide asymptotically
tight bounds in the Hermitian case on
the typical value of the second largest
eigenvalue.  The key idea is the use of Schwinger-Dyson equations from
lattice gauge theory to efficiently compute averages over the unitary group.
\end{abstract}
\maketitle

Recently, two papers\cite{em1,em2} introduced the idea of {\it
expander maps}: quantum analogues of expander
graphs.  An expander graph\cite{hlw} may be defined in several ways.  One
is the property of
having a large number of vertices, a small coordination number for
each vertex, and also having a gap in the spectrum of the diffusion equation
on the graph, so that a particle classically diffusing on an expander graph
rapidly loses memory of where it started.
In the quantum case, we replace the random process of diffusion by
a completely positive, trace preserving map
${\cal E}(M)$.  We define a quantum expander to be such
a map from the space of $N$-by-$N$ matrices $M$ to the same
space with the following properties.  First, $N$ is large, in analogy
to the large number of vertices.  Second,
the map has
eigenvalues $\lambda_1,\lambda_2,\lambda_3,...,\lambda_{N^2}$,
with $\lambda_1=1$ and
$|\lambda_a|\leq 1-\delta$ for all $a>1$ so that the eigenvalue spectrum
has a gap.
Finally, the map can be written as
\be
\label{mapdefn}
{\cal E}(M)=\sum_{s=1}^D A^{\dagger}(s) M A(s)
\ee
for some relatively small value of $D$, with $\sum_{s=1}^D A(s) A^{\dagger}(s)=
\openone$ so that the map is trace preserving, and with
$\sum_{s=1}^D A^{\dagger}(s) A(s)=\openone$, so that
the eigenvector corresponding to eigenvalue $\lambda_1$ is 
$(1/\sqrt{N})\openone$.  Here, $\openone$ is the $N$-by-$N$ unit matrix.
This requirement of small $D$ is in analogy to the low coordination number.

These maps were applied in \cite{em1} to construct many-body states in one
dimension with the property of having a short correlation length
(this corresponds to the gap $\delta$
in the spectrum of eigenvalues of ${\cal E}$),
small Hilbert space dimension on each site (this corresponds to the small $D$),
and yet large entanglement entropy (this corresponds to the large
entropy of the eigenvector of ${\cal E}$ with unit eigenvalue).  Since
expander graphs have a large number of applications in problems dealing
with classical statistics, 
such as in error-correcting codes\cite{ecc},
derandomization,
and the PCP theorem\cite{pcp},
to name a few, it seems worth further exploring the quantum case.

A number of possible forms of an expander map are possible:
in \cite{em2} an expander
was defined as having 
\be
\label{unitary}
A(s)=\frac{1}{\sqrt{D}} U(s),
\ee
for some unitary matrices $U(s)$, and in fact this is the form
of $A(s)$ considered in this paper.  However, the more general
definition with arbitrary $A(s)$ constrained by 
$\sum_{s=1}^D A^{\dagger}(s) A(s)=\openone,
\sum_{s=1}^D A(s) A^{\dagger}(s)=\openone$ seems also useful; in fact,
although we do not consider it in this paper, it may be useful to weaken
this constraint further, and explore
the properties of completely positive, trace preserving maps, with no
other constraint on the $A$, requiring only that the entropy of the
density matrix which is the eigenvector with unit eigenvalue is 
large\cite{formal}.

Our goal is to try to find families of maps with arbitrarily large $N$,
such that the gap $\delta$ is bounded below by some $N$-independent constant and
such that $D$ does not grow too rapidly with $N$.
The first paper\cite{em1} provided an
explicit construction of such a family of maps with $D$ of order $\log(N)$
and provided numerical evidence for an alternate construction with
$D$ {\it independent} of $N$.  The second paper\cite{em2} gave yet a different
construction with $D$ of order $\log(N)$ but also provided a construction
that had $D$ independent of $N$ and succeeded in proving an $N$-independent
lower bound on the gap $\delta$ in this case.

Experience with expander graphs suggests that, while finding deterministic
constructions of them is difficult\cite{det},
with high probability a
{\it random} graph of fixed coordination number greater than 2 is
an expander\cite{rand}.  Thus, the natural question is to investigate whether
Eqs.~(\ref{mapdefn},\ref{unitary}) will give an expander map
if the matrices $U$ are chosen randomly from the unitary group $U(N)$ using
the Haar measure.  We consider
two cases.  In the first case, the map ${\cal E}$ is non-Hermitian and the
$D$ matrices are chosen independently at random.  In the second
case the matrices $U(s)$ are chosen independently at random for $s=1...D/2$ and
we pick $U(s+D/2)=U(s)^{\dagger}$.  In this case, $D$ is
even, and the map ${\cal E}$
is Hermitian and has real eigenvalues.
In this paper we begin in generality with
the non-Hermitian case, but then restrict to the Hermitian case
for simplicity of notation.

In the Hermitian case, we consider $D\geq 4$, while in the non-Hermitian
case we consider $D\geq 2$, as otherwise we would clearly not have
an expander.
Let $\lambda_2$ be the eigenvalue with the second largest absolute value
of all eigenvalues other than $\lambda_1$.
Let
\be
\lambda_{H}=\frac{2\sqrt{D-1}}{D}.
\ee

The main result of this paper is
that, in the Hermitian case, for any $\epsilon>0$ the probability that
$|\lambda_2|$ is within $\epsilon$ of $\lambda_{H}$ approaches
unity as $N\rightarrow \infty$.  Interestingly, this is the same
as the recently proven tight bound\cite{fried} in the classical
case, but the proof in the
quantum case is much simpler.

The proof is based on a version of the trace method.  We begin by
introducing the trace method and describing its application to the Hermitian
and non-Hermitian cases.  We then give lower bounds on
$|\lambda_2|$ based on the return probability of a random walk on a Cayley
tree and discuss some numerical results.  We next
introduce a set of Schwinger-Dyson equations, analogous to those used
in lattice gauge theory\cite{ek}.  This is the key step which enables us
to take averages over the unitary group efficiently.  We will use these
equations to develop a {\it convergent}
perturbation theory in $1/N$ for various traces of
unitary matrices, and bound the correction terms in this perturbation theory.
We start with a loose bound, giving a loose bound on $|\lambda_2|$, and
then tighten to get the tight bound above.  Finally, in an appendix
we discuss a related problem of ``quantum edge expanders", which gives
an analogue in the quantum case of the combinatorial definition of a classical
expander graph.

The space of $N$-by-$N$ complex matrices $M$ has a natural inner
product: $(M,N)={\rm tr}(M^{\dagger}N)$.
With respect to this inner product, an orthonormal basis of matrices
consists of the matrices $M(i,j)$, defined to have a $1$ in the $i$-th
row and $j$-th column, and zeroes everywhere else.  Given
this inner product, we can consider the space of $N$-by-$N$ matrices
as an $N^2$-dimensional vector space, with ${\cal E}$ acting as
a linear operator on this space.  Then, in the Hermitian case,
it is possible to find a linear operator $V$, which is unitary
with respect to this inner product, such that ${\cal E}=V^{\dagger} \Lambda  V$,
with $\Lambda$ being a diagonal matrix with entries $\lambda_a$.  Note
that here ${\cal E}, V$, and $\Lambda$ are all $N^2$-by-$N^2$ dimensional
matrices.
In the non-Hermitian
case, we can write ${\cal E}=V^{\dagger} T V$, with $T$ an upper triangular
matrix whose diagonal entries are the eigenvalues $\lambda_a$.
Thus,
\begin{eqnarray}
\label{tr}
\sum_{i,j} \Bigl({\cal E}^m(M(i,j)),{\cal E}^m(M(i,j))\Bigr)
&=&\sum_{i,j}\Bigl(T^m(M(i,j), T^m(M(i,j))\Bigr)
\\ \nonumber
&\geq &
\sum_{a=1}^{N^2} |\lambda_a|^{2m},
\end{eqnarray}
where ${\cal E}^m(M)$ denotes acting with the map ${\cal E}$ successively
$m$ times on $M$, and similarly for $T^m(M)$.
In the case where ${\cal E}$ is Hermitian, Eq.~(\ref{tr}) is
an equality.  

To simplify notation, we now restrict to the Hermitian case.
In this case, Eq.~(\ref{tr}) can be replaced by
\be
\label{trH}
\sum_{i,j} \Bigl(M(i,j),{\cal E}^m(M(i,j))\Bigr)
=\sum_{a=1}^{N^2} |\lambda_a|^{m}\geq 1+|\lambda_2|^m,
\ee
where we pick $m$ to be an even integer.
Then,
\begin{eqnarray}
\label{e0for1}
\sum_{a=1}^{N^2} |\lambda_a|^m&= &
\sum_{i,j} \Bigl(M(i,j),{\cal E}^m(M(i,j))\Bigr)
\\ \nonumber
&= &
\Bigl(\frac{1}{D}\Bigr)^m \sum_{s_1=1}^D
\sum_{s_2=1}^D...
\sum_{s_{m}=1}^D
{\rm tr}(U(s_m+D/2) ...
U(s_2+D/2) U(s_1+D/2))
{\rm tr}(U(s_1) U(s_2) ... U(s_m))].
\end{eqnarray}
For notational
convenience, we identify $s_i+D$ with $s_i$ throughout this paper, so that $s_i$
is a periodic variable with period $D$.

Averaging $U(1),...,U(D)$ over the unitary group we find that
$E[\sum_{i,j} \Bigl(M(i,j),{\cal E}^m(M(i,j))\Bigr)]=
E[\sum_{a=1}^{N^2} |\lambda_a|^m]$,
where $E[...]$ denotes the given average.
Averaging Eq.~(\ref{e0for1}) we find
\be
\label{av}
E_1\equiv \Bigl(\frac{1}{D}\Bigr)^m \sum_{s_1=1}^D
\sum_{s_2=1}^D...
\sum_{s_{m}=1}^D
E_0(s_1,...,s_m)
=E[\sum_{a=1}^{N^2} |\lambda_a|^m],
\ee
where
\begin{eqnarray}
\label{e0def}
E_0(s_1,...,s_m) &\equiv &
E[
{\rm tr}(U^{\dagger}(s_m) ...
U^{\dagger}(s_2) U^{\dagger}(s_1))
{\rm tr}(U(s_1) U(s_2) ... U(s_m))]\\ \nonumber &=&
E[
{\rm tr}(U(s_m+D/2) ...
U(s_2+D/2) U(s_1+D/2))
{\rm tr}(U(s_1) U(s_2) ... U(s_m))].
\end{eqnarray}

\section{Lower Bounds and Numerical Results}

In this section we present lower bounds 
on $|\lambda_2|$ based on random walks on a Cayley tree and
then provide some numerical results.  In the Hermitian
case, it is possible, for certain choices of $s_1,...,s_m$ in
either Eq.~(\ref{av}) or Eq.~(\ref{e0for1}), that
the trace ${\rm tr}(U(s_1) U(s_2) ... U(s_m))$ can be reduced to
a trivial trace of the identity matrix by canceling successive
appearances of $U(s) U(s+D/2)$ and replacing them with $\openone$.
The contribution of such choices to $E_1$ is
proportional to a return probability of a random walk on a Cayley tree as will
be seen.

We begin with an upper bound on the number of such choices:
consider the unitaries $U(s_1) U(s_2) ... U(s_k)$ for some $k$, $0\leq
k \leq m$.  After making all possible cancellations of successive terms,
$U(s) U(s+D/2)$, this sequence of unitaries may be reduced to another
sequence of unitaries $U(s'_1(k)) U(s'_2(k)) ... U(s'_{l(k)}(k))$,
for some $l(k)\leq k$.
Then, consider the sequences of unitaries $U(s_1) U(s_2) ... U(s_{k+1})$.
After making the same cancellations, and then possibly canceling
$U(s_{k+1})$ against $U(s'_{l(k)}(k))$, we find a new sequence of
unitaries, $U(s'_1(k+1)) U(s'_2(k+1) ... U(s'_{l(k+1)}(k+1))$
with $l(k+1)=l(k)\pm 1$
and $s'_j(k+1)=s'_j(k)$ for $j<l(k)$.
When $l(k+1)=l(k)-1$, then $s_{k+1}$
is determined by $s_k$.   When $l(k+1)=l(k)+1$, then if $l(k)>0$ there are
$D-1$ possible values of $s_{k+1}$, while if $l(k)=0$ there are $D$
possible values.  Note that $l(k)\geq 0$ for all $k$.
We define $N(l(m),m)$ to be the number of choices of $s_1,s_2,...,s_m$ which
give rise to the given $l(m)$.  This is precisely the number of
random walks of length $m$, on a tree with $D$ daughters at the root and
$D-1$ daughters for every other node, that end at a distance $l(m)$ from
the root.  Note that $N(0,m)$ is equal to $D^m$ times the return probability
of a random walk of length $m$ on the Cayley tree.

An upper bound on $N(0,m)$ is given by
\be
\label{N0m}
N(0,m) \leq (D-1)^{m/2} \frac{m!}{(m/2)!(m/2)!} \leq (D-1)^{m/2} 2^m.
\ee
To show Eq.~(\ref{N0m}), we consider a related problem:
consider sequences of $l(k)$ in which $l(k)$ may become {\it negative}, while
the number of choices of $s_m$ is considered to be $D-1$ whenever
$l(k+1)=l(k)+1$, and the number of choices is considered to $1$ whenever
$l(k)=l(k)-1$.  This give an overestimate of the number of sequences, and
gives the value in Eq.~(\ref{N0m}).

On the other hand, a lower bound on $N(0,m)$ is given by assuming
that if $l(k+1)=l(k)+1$ there are only $D-1$ possible choices of $s_{k+1}$,
regardless of $l(k)$, in which case we find that
\be
\label{n0below}
N(0,m) \geq c \times (2\sqrt{D-1})^m/(m+1)^{3/2},
\ee
for some constant $c$ of order unity.  These bounds, (\ref{N0m},\ref{n0below}),
are compeletely standard bounds\cite{hlw}, and we only repeat their derivation
for completeness.

We now use Eq.~(\ref{n0below})
to
get a lower
bound on $|\lambda_{2}|$ for any completely positive map where the matrices
$A(s)$ are given by Eq.~(\ref{unitary}).  We emphasize that, while the
upper bounds elsewhere in this paper are upper bounds on the typical behavior
of $|\lambda_2|$, the present result is valid for any such map given
by Eqs.~(\ref{mapdefn},\ref{unitary}), and is a quantum analogue of
the Alon-Boppana bound\cite{ab}.
First, $\sum_{a=1}^{N^2} |\lambda_a|^m
= 1+ \sum_{a=2}^{N^2} |\lambda_a|^m \leq 1+(N^2-1) |\lambda_2|^m <
1+N^2 |\lambda_2|^m$.
Note that the product of traces in Eq.~(\ref{e0for1}) is equal to
$|{\rm tr}(U(s_1) U(s_2) ... U(s_m))|^2$ and so is
positive for all choices of $s_1,...,s_m$. 
If $l(m)=0$, then 
$|{\rm tr}(U(s_1) U(s_2) ... U(s_m))|^2=N^2$, and so
the contribution of terms
with $l(m)=0$ to the sum in Eq.~(\ref{e0for1}) is equal to
$N^2 N(0,m)/D^m$.
Therefore,
\be
1+N^2 |\lambda_2|^m \geq \sum_{a=1}^{N^2} |\lambda_a|^m \geq
N^2 \frac{N(0,m)}{D^m}\geq c N^2 \lambda_{H}^m/m^{3/2}.
\ee
Thus, $|\lambda_2| \geq \lambda_{H} (c/m^{3/2})^{1/m}
[1-m^{3/2}/(c \lambda_H^m N^2)]^{1/m}$.
Picking $m=[\ln(cN^2/2)-(3/2)\ln(\ln(cN^2/2))]/\ln(1/\lambda_{H})$, we find
\be
\label{lowB}
|\lambda_2| \geq \lambda_{H} (1-{\cal O}(\ln(\ln(N))/\ln(N))).
\ee

\begin{figure}[tb]
\centerline{
\includegraphics[scale=0.4]{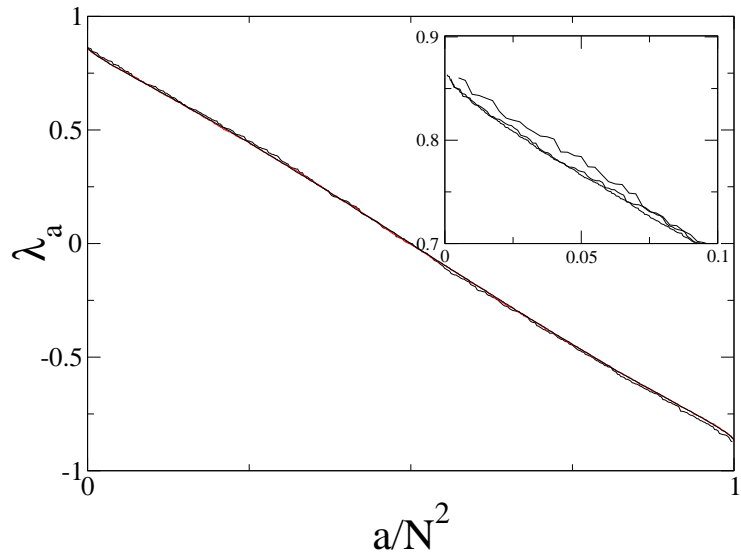}}
\caption{Eigenvalues from numerical diagonalization of a completely positive
map based on the construction in \cite{em1} using expander graphs, for
$N=20,30,50$.
The eigenvalue with eigenvalue unity
is not shown.  The second largest eigenvalue is at roughly
$\sqrt{3}/2$.  Only a single realization is shown for each $N$.  The inset shows
a detail of the behavior at small $a$.  Curves in the inset are $N=20,30,50$
from top to bottom; the curves in the main figure are not distinguishable.}
\end{figure}

A very interesting question is to see whether a bound such as
(\ref{lowB}) still holds for arbitrary trace preserving,
completely positive Hermitian maps ${\cal E}(M)$.  As a partial step towards
this more general bound, note that the bound of Eq.~(\ref{lowB}) can be
readily generalized to the following case: let $A(s)=\sqrt{P(s)} U(s)$,
with the numbers $P(s)$ obeying $\sum_{s=1}^D P(s)=1$, and with
$U(s)=U(s+D/2)^{\dagger}$ and $P(s)=P(s+D/2)$.  Eq.~(\ref{unitary})
is a special case of this with $P(s)=1/D$.

As stated before,
the main result of this paper is that, for any $\epsilon$, when the unitary
matrices are chosen randomly, the probability
that $|\lambda_2|$ is within $\epsilon$ of $\lambda_H$ approaches
unity when $N$ becomes large.  Interestingly, this seems to be true in more
generality than just for unitary matrices chosen with the Haar measure.
Using the construction in \cite{em1}, in which we pick a random graph with
constant coordination number and derive unitary matrices from that graph
and from certain random phases,
numerical studies also show that $|\lambda_2|$ is close to $\lambda_H$.  We
show in Fig.~1 the results of numerical diagonalization of systems
of size $N=20,30,50$, so that there are $400,900,2500$ eigenvalues respectively.
The second largest eigenvalue is indeed very close
to $\sqrt{3}/2$.  After sorting the eigenvalues by $\lambda_a$, from most
positive to most negative, we plotted the eigenvalues as a function of
$a/N^2$: the scaling collapse of the curves is extremely good.

\section{Bounds on Eigenvalues}
\subsection{Schwinger-Dyson Equations}
We will develop a perturbation theory in $1/N$ to estimate the
average (\ref{av}), which is a product of two traces.
To do this, we will develop general machinery for computing the average
over the unitary group of products of an arbitrary number of traces.
Consider such a product of the form:
\be
\label{prodtrace}
E[L_1 L_2 ... L_k],
\ee
where
\be
L_1={\rm tr}(U(s_{1,1})U(s_{1,2})...U(s_{1,m_1})), \quad
L_2={\rm tr}(U(s_{2,1})U(s_{2,2})...U(s_{2,m_2})), \quad...
\ee
Here we have an average of $k$ traces, $L_1,...L_k$,
each of which is a product of
$m_k$ unitary matrices.
We now present the Schwinger-Dyson equations.

Let $T^a$, for $a=1...N^2$, be Hermitian matrices such that
\be
\sum_{a=1}^{N^2} T^a_{\mu\nu} T^a_{\rho\sigma}=\delta_{\mu\sigma}\delta_{\nu\rho}.
\ee
Then
\be
\sum_{a=1}^{N^2} (T^a T^a)_{\mu\nu}=N \delta_{\mu\nu}.
\ee
To compute the average in Eq.~(\ref{prodtrace}), we begin with
\be
\label{pt2}
E[
{\rm tr}(T^a U(s_{1,1})U(s_{1,2})...U(s_{1,m_1}))
L_2 ... L_k].
\ee
We then use the invariance of the average over the unitary group under
an infinitesimal change in variables:
\begin{eqnarray}
\label{change}
U(s_1)\rightarrow (1+i\epsilon T^a)U(s_1) \\ \nonumber
U(s_1+D/2)\rightarrow U(s_1+D/2)(1-i\epsilon T^a),
\end{eqnarray}
where we recall that $U(s+D/2)=U(s)^{\dagger}$.
Applying the change in variables given in Eq.~(\ref{change}) to
Eq.~(\ref{pt2}), and summing over $a$ and dividing by $N$, we find that:
\begin{eqnarray}
\label{SD}
&&
E[
{\rm tr}(U(s_{1,1})U(s_{1,2})...U(s_{1,m_1}))
L_2 ... L_k]\\ \nonumber
&&=
-
\frac{1}{N}
\sum_{j=2}^{m_1} \delta_{s_{1,1},s_{1,j}}
E[
{\rm tr}(U(s_{1,1})...U(s_{1,j-1}))
{\rm tr}(U(s_{1,j})...U(s_{1,m_1}))
L_2 ... L_k]\\ \nonumber
&&+
\frac{1}{N}
\sum_{j=2}^{m_1} \delta_{s_{1,1},s_{1,j+D/2}}
E[
{\rm tr}(U(s_{1,1})...U(s_{1,j}))
{\rm tr}(U(s_{j+1,1})...U(s_{1,m_1}))
L_2 ... L_k]\\ \nonumber
&&-
\frac{1}{N}
\sum_{l=2}^k
\sum_{j=1}^{m_l}
\delta_{s_{1,1},s_{l,j}}
E[
{\rm tr}(U(s_{1,1})...U(s_{1,m_1}) U(s_{l,j}) U(s_{l,j+1}) ...
U(s_{l,j-1}))
L_2 ... L_{l-1} L_{l+1} ... L_k] \\ \nonumber
&&+
\frac{1}{N}
\sum_{l=2}^k
\sum_{j=1}^{m_l}
\delta_{s_{1,1},s_{l,j}+D/2}
E[
{\rm tr}(U(s_{1,1})...U(s_{1,m_1}) 
U(s_{l,j+1}) U(s_{l,j+2}) ...
U(s_{l,j-1}) U(s_{l,j}))
L_2 ... L_{l-1} L_{l+1} ... L_k].
\end{eqnarray}
We simplify the second and fourth lines after the equality sign
of the above equation using $U(s) U(s+D/2)=\openone$ to get
\begin{eqnarray}
\label{SD2}
&&
E[
{\rm tr}(U(s_{1,1})U(s_{1,2})...U(s_{1,m_1}))
L_2 ... L_k]\\ \nonumber
&&=
-
\frac{1}{N}
\sum_{j=2}^{m_1} \delta_{s_{1,1},s_{1,j}}
E[
{\rm tr}(U(s_{1,1})...U(s_{1,j-1}))
{\rm tr}(U(s_{1,j})...U(s_{1,m_1}))
L_2 ... L_k]\\ \nonumber
&&+
\frac{1}{N}
\sum_{j=2}^{m_1} \delta_{s_{1,1},s_{1,j+D/2}}
E[
{\rm tr}(U(s_{1,2})...U(s_{1,j-1}))
{\rm tr}(U(s_{j+1,1})...U(s_{1,m_1}))
L_2 ... L_k]\\ \nonumber
&&-
\frac{1}{N}
\sum_{l=2}^k
\sum_{j=1}^{m_l}
\delta_{s_{1,1},s_{l,j}}
E[
{\rm tr}(U(s_{1,1})...U(s_{1,m_1}) U(s_{l,j}) U(s_{l,j+1}) ...
U(s_{l,j-1}))
L_2 ... L_{l-1} L_{l+1} ... L_k] \\ \nonumber
&&+
\frac{1}{N}
\sum_{l=2}^k
\sum_{j=1}^{m_l}
\delta_{s_{1,1},s_{l,j}+D/2}
E[
{\rm tr}(U(s_{1,2})...U(s_{1,m_1}) 
U(s_{l,j+1}) U(s_{l,j+2}) ...
U(s_{l,j-1}))
L_2 ... L_{l-1} L_{l+1} ... L_k].
\end{eqnarray}
These Schwinger-Dyson equations are quite long when written out,
but in reality are quite simple.  Let us apply them to compute the
average of ${\rm tr}(U){\rm tr}(U^{\dagger})$ over
unitary matrices $U$.  We find after one iteration of Eq.~(\ref{SD2})
that this is equal to
$(1/N) {\rm tr}(\openone)=1$.
Now consider a more complicated example, to compute the
average of ${\rm tr}(U U){\rm tr}(U^{\dagger} U^{\dagger})$ over
unitary matrices $U$.   Then, the Schwinger-Dyson equations give after
the first iteration:
$(1/N)(-E[{\rm tr}(U) {\rm tr}(U) {\rm tr}(U^{\dagger} U^{\dagger}]
+2 E[{\rm tr}(U U^{\dagger}])=
-(1/N)E[{\rm tr}(U) {\rm tr}(U) {\rm tr}(U^{\dagger} U^{\dagger}]
+2$.  We then apply the equations again to the average
$E[{\rm tr}(U) {\rm tr}(U) {\rm tr}(U^{\dagger} U^{\dagger}]$, giving
$E[{\rm tr}(U) {\rm tr}(U) {\rm tr}(U^{\dagger} U^{\dagger}]=
(1/N)(-E[{\rm tr}(U U) {\rm tr}(U^{\dagger} U^{\dagger})]
+2E[{\rm tr}(U) {\rm tr}(U^{\dagger}])$.  Since we have already worked out
$E[{\rm tr}(U) {\rm tr}(U^{\dagger}])=1$, we have
$E[{\rm tr}(U) {\rm tr}(U) {\rm tr}(U^{\dagger} U^{\dagger}]=
-(1/N)E[{\rm tr}(U U) {\rm tr}(U^{\dagger} U^{\dagger})]+2$.
Thus, putting it all together, we find that
\be
\label{example}
E[{\rm tr}(U U){\rm tr}(U^{\dagger} U^{\dagger})]=
2+(1/N^2)
E[{\rm tr}(U U){\rm tr}(U^{\dagger} U^{\dagger})]-(2/N^2),
\ee
and hence
$E[{\rm tr}(U U){\rm tr}(U^{\dagger} U^{\dagger})]=2$.

We now describe the general algorithm for reducing traces of the
form (\ref{prodtrace}).  We initially cancel all pairs of
matrices $U(s)U(s+D/2)$ appearing successively in the same trace,
replacing them with $\openone$.
We then apply the equation (\ref{SD2}).  Then, we cancel all
pairs of matrices
$U(s)U(s+D/2)$ appearing successively
in the resulting traces, replace the trace ${\rm tr}(\openone)$ by
$N$, and repeat this procedure for each term.
After the first application
of Eq.~(\ref{SD2}), the number of terms on the right-hand side will
be at most $m_{total}-1$, where $m_{total}\equiv m_1+m_2+...+m_k$.
Applying the equations repeatedly will
generate more and more terms at each application.  We regard this as
a branching process: each term on the right-hand side can be then fed
back into the left-hand side of the equation to generate new terms on the
right-hand side.  Note that at every stage, each term will produce at
most $m_{total}-1$ terms on the right-hand side since the total number
of unitary matrices which appear in the traces, $m_1+m_2+...+m_k$, will
always be at most $m_{total}$.  If
the number of unitary matrices becomes equal to zero in some term after
$n$ iterations of Eq.~(\ref{SD2}),
then we are left with only trivial traces and we say that this term
``terminates" at level $n$.

This algorithm generates an infinite series, where the $n$-th term
in the series is equal to the sum of all terms terminating at the $n$-th
level.
We claim (and will show later when we discuss
the convergence of the series) that, if $m_{total}\leq N$, then
this series is absolutely
convergent and also the average of the original
trace is equal to the sum over all levels, $n\geq 1$, of the terms which
terminate at each level, so that the series converges to the desired
answer.

This series is in fact an infinite series for many simple examples.
In fact, for
$E[{\rm tr}(U U){\rm tr}(U^{\dagger} U^{\dagger})]$ we find that
after two repetitions of the process, the same average
$E[{\rm tr}(U U){\rm tr}(U^{\dagger} U^{\dagger})]$ has re-appeared,
as can be seen on the right-hand side of
Eq.~(\ref{example}), and thus the algorithm above does not
ever finish because there are always some terms with nontrivial traces.
In this particular case, however, although the algorithm does
not ever finish,
the sum of the terms terminating at any level $n>1$
is equal to zero; 
in other cases\cite{example}
this is not true and the given series has an infinite
number of nonvanishing coefficients.   
We will later see how this infinite series is related to
an infinite series in $1/N$ for the given trace.

We will apply this procedure to the trace $E_0=E[
{\rm tr}(U(s_m+D/2) ...
U(s_2+D/2) U(s_1+D/2))
{\rm tr}(U(s_1) U(s_2) ... U(s_m))]$.  Thus,
$L_1=
{\rm tr}(U(s_m+D/2) ...
U(s_2+D/2) U(s_1+D/2))$, and $L_2=
{\rm tr}(U(s_1) U(s_2) ... U(s_m))]$.  Begin by reducing all successive
pairs of a unitary matrix followed by its Hermitian conjugate.  What is left is
two traces $L_1,L_2$ such that $m_1=m_2$ and
$s_{1,i}=s_{2,m_2+1-i}+D/2$.  We will proceed by estimating the probability
of different values of $m_1$ given a random choices of $s_1,...,s_m$, and
then estimating the behavior of $E_0$ for the given $m_1=m_2$.

\subsection{Length of the Reduced Trace}
In this subsection, we will estimate the number of choices of
$s_1,...,s_m$ such that
the reduced traces, $L_1,L_2$ have a given
$m_1=m_2$. 

We start with the case $m_1=m_2=0$ in which case $E_0=N^2$.
The number of different choices of $s_1,...,s_m$ with $m_1=m_2=0$ is
given by Eq.~(\ref{N0m})
so the contribution of all such choices to $E_1$ is bounded by
\be
\label{ctrepprob}
N^2 D^{-m} (D-1)^{m/2} 2^m = N^2 \lambda_H^m.
\ee

We can also bound the number of choices of $s_1,...,s_m$ which give
a given $m_1>0$.  In this case, $l(m)=m_1$ and $s'_{j}(m)=s_{1,j}$.
Using the same argument as gave Eq.~(\ref{N0m}), the number of such choices
is bounded by
\be
\label{nm1}
(D-1)^{m_1/2} (D-1)^{m/2} 2^m.
\ee
This number is independent of the particular values of
$s_{1,1},...,s_{1,m_1}$.  There are $[D/(D-1)] (D-1)^{m_1}$ different
possible values of $s_{1,1},...,s_{1,m_1}$
and therefore the total number of choices
of $s_1,...,s_m$ which give rise to a given choice of
$s_{1,1},...,s_{1,m_1}$ is bounded by
\be
\label{nchoice}
\frac{D-1}{D} \Bigl(\frac{1}{\sqrt{D-1}}\Bigr)^{m_1} (D-1)^{m/2} 2^m.
\ee

\subsection{Nontrivial Words}
We now consider the case the $m_1>0$.
After the first application of Eq.~(\ref{SD2}), the term on the fourth line
with $l=2$ and $j=m$ reduces the trace to
$(1/N)E[{\rm tr}(U(s_{1,2})...U(s_{1,m_1})U(s_{2,1})...U(s_{2,m_1-1}))]=(1/N)E[\openone]=1$,
so that the series terminates at level $n=1$.
There may also be other terms which reduce the trace to a trival
one after a single application of Eq.~(\ref{SD2}) if the sequence of
values $s_{1,1}s_{1,2}...s_{1,m_1}$ has a symmetry under a shift:
$s_{j,1}=s_{j+m_1/o, 1}$ for some $o>1$ which we refer to as
the period of the shift.  Here, we treat the
index $j$ as periodic with period $m_1$.  For example, the problem
studied in Eq.~(\ref{example}) has such a symmetry under a shift with $o=2$.
In the event that there is such a shift symmetry, then the sum of
terms terminating at level $n=1$ is equal to $o$.  For a given $m_1$,
the number of choices of $s_{1,1},...,s_{1,m_1}$ which
have a shift symmetry with period $o$ is bounded by
$[D/(D-1)] (D-1)^{m_1/o}$.  
Thus, from Eq.~(\ref{nchoice}), the total number of choices
of $s_1,...,s_m$ which give rise to a given $m_1,o$ is bounded by
$(D-1)^{m_1/o} \Bigl(\frac{1}{\sqrt{D-1}}\Bigr)^{m_1} (D-1)^{m/2} 2^m$.
Thus, the contribution to $E_1$
of terms terminating at level $n=1$
is bounded by
\be
\label{prlm}
1+\sum_{m_1\leq m}
\sum_{o=2}^{m_1} 
o 
(D-1)^{m_1/o} \Bigl(\frac{1}{\sqrt{D-1}}\Bigr)^{m_1} \lambda_H^m.
\ee
The term in Eq.~(\ref{prlm}) with $o=2$ is is bounded by
$2 m \lambda_H^m$, while the sum of terms in Eq.~(\ref{prlm}) with
$o>2$ is bounded by $(D-1)^{-1/6}/[1-(D-1)^{-1/6}]^2 \lambda_H^m$, which
for $D\geq 4$ is bounded by $30 \lambda_H^m$ so that Eq.~(\ref{prlm}) is
bounded by
\be
\label{leveloneb}
1+2 m \lambda_H^m+30 \lambda_H^m.
\ee

We will now bound the sum of all terms terminating at
level $n>1$.
Assuming that the sequence $s_{1,2}...s_{1,m_1}$ lacks the shift symmetry
discussed above, this is the only term which terminates at level 1, and
the other terms which appear after the first iteration do not terminate
and continue to branch, but some of their descendents will terminate
at lower levels.

We can estimate the value of a term which terminates at a given level $n>1$
as follows.  First, there is a sign equal to plus or minus $1$.  Next,
there is a factor of $(1/N)^n$.  Finally, there is a factor of $N$ for
each trace of the form ${\rm tr}(\openone)$ that appeared in this process.
Suppose there are $p$ such traces, giving a factor of $N^p$.  How big
can $p$ be?  Initially we have $k=2$ different traces.  The given term
at level $n$ arose from a specific choice of terms on the right-hand side
of Eq.~(\ref{SD2}) on the first iteration.  This specific choice has $k_1$
different traces in it, with $k_1$ equal to either $1$ or $3$.
After the second iteration there are $k_2$ traces, then $k_3$, and so on.
The number of traces
$k_2,k_3,...$ can be determined as follows: an application of
Eq.~(\ref{SD2}) may increase the number of traces by one if the term
arises from the first or second line on the right-hand side, or may decrease
the number of traces by one if the term arises from the third or fourth
line on the right-hand side of Eq.~(\ref{SD2}).  Next, some of
the traces may be trivial.  In the event that the term arose
from the first, second, or third line of Eq.~(\ref{SD2}) it is not possible for
any of the traces to be trivial, under the assumption that any repetitions
of the form $U(s)U(s+D/2)$ have been replaced by $\openone$ in the trace.
However, in the event that the term arose from the fourth line,
then it is possible for one of the traces to be trivial, increasing
$p$ by one.  Thus, for each $b\leq n$,
$k_b-k_{b-1}$ is equal to either $+1,-1,$ or $-2$.
Let $q$ be equal to the number of times the first or second line was used
from Eq.~(\ref{SD2}) and $n-q$ equal the number of times the third or fourth
line was used.  Then, in order for all traces to be trivial in this particular
term resulting from $n$ iterations of Eq.~(\ref{SD2}),
\be
2+q-(n-q)-p=0.
\ee
Also, since $p$ can only increase when a term from the fourth line is
used,
\be
p \leq n-q.
\ee
Thus,
\be
p \leq \lfloor (2+n)/3 \rfloor.
\ee
Therefore, the value of a term terminating
at the $n$-th level, $n>0$, is bounded in absolute
value by
\be
\label{valBd}
N^{\lfloor (2+n)/3\rfloor -n}.
\ee

The number of terms terminating
at the $n$-th level is bounded by
\be
\label{numBd}
(2m-1)^n.
\ee
Note also that there are no terms terminating at level $n=2$: if the
term does not terminate at level $1$, then there are either $1$ or
$3$ traces after the first iteration of Eq.~(\ref{SD2}), and then
there is no way to have the term terminate at level $2$.
Thus, the sum of terms terminating at level
$n>1$ is bounded in absolute value by
\begin{eqnarray}
\label{bndSum}
&&(2m)^3 N^{-2}+(2m)^4 N^{-2}+(2m)^5 N^{-3}+(2m)^6 N^{-4}+(2m)^7 N^{-5}
+...
\\ \nonumber
&\leq &
8m^3 N^{-2}+16m^4 N^{-2}\Bigl(1+2m N^{-1}+4m^2 N^{-2}\Bigr)
\frac{1}{1-8m^3 N^{-2}}.
\end{eqnarray}

\subsection{Convergence of Series}
We now show the claim that, for $m_{total}\equiv m_1+m_2+...+m_k \leq N$, the average
$E[L_1 L_2 ... L_k]$
is indeed equal to
the sum over all levels $n\geq 1$ of the number of terms terminating
at each level and that the series is absolutely convergent.
After $n$ iterations of Eq.~(\ref{SD2}) some of the
terms have terminated.  There are at most $(m_{total}-1)^n$
terms which have not
terminated, since there are at most $(m_{total}-1)^n$ terms.  Each of these
terms is equal to plus or minus one times $N^{-n}$ times $N^{p_n}$
where $p_n$ is the number of times a trivial trace appeared in
the process, times
the average of a product of traces.
There are at most $m_{total}-p_n$
different traces in the product since there were originally at most
$m_{total}$ unitary matrices.  
Thus, since each trace is bounded in absolute
magnitude by $N$, the sum of all terms which have not terminated after
$n$ applications of Eq.~(\ref{SD2}) is bounded in absolute value by
$(m_{total}-1)^n N^{-n} N^{m_{total}}$ which converges to zero as $n\rightarrow \infty$
for $m_{total}\leq N$.  Thus, the difference between the sum of the terms
terminating at the
first $n$ levels and the actual value of the average $E[L_1 L_2 ... L_k]$ converges to
zero as $n \rightarrow \infty$.
The sum of all terms terminating at a given
level is bounded in absolute value by the number of such terms, times
$N^{-n} N^{p_n}$, and so is bounded by $(m_{total}-1)^n N^{-n} N^{m_{total}}$ and
so the series is absolutely convergent for $m_{total}\leq N$.
This shows the desired claim.

\subsection{Loose Bound}
Adding the results in Eq.~(\ref{ctrepprob},\ref{leveloneb},\ref{bndSum}),
we find that for $2m<N$, 
\be
\label{looseB}
E_1\leq
1+[N^2+2 m + 30] \lambda_H^m
+8m^3 N^{-2}+16m^4 N^{-2}\Bigl(1+2m N^{-1}+4m^2 N^{-2}\Bigr)
\frac{1}{1-8m^3 N^{-2}}.
\ee

We now pick $m=\log(N^4)/\log(1/\lambda_H)$, so
\be
E_1\leq 1+16 (1+{\rm o}(1)) [\log(N^4)/\log(1/\lambda_H)]^4 N^{-2},
\ee
where ${\rm o}(1)$ denotes terms asymptotically tending to zero as
$N\rightarrow \infty$.
Thus, the average of $|\lambda_2|$ over the unitary group is bounded by
\begin{eqnarray}
&& \Bigl\{16 (1+{\rm o}(1)) [\log(N^4)/\log(1/\lambda_H)]^4 N^{-2}\Bigr\}^{
\log(1/\lambda_H)/\log(N^4)}
\\ \nonumber
&=&(1+{\rm o}(1)) \lambda_{loose}(D),
\end{eqnarray}
where
\be
\lambda_{loose}(D) \equiv \sqrt{\lambda_H}=
\sqrt{\frac{2\sqrt{D-1}}{D}}.
\ee
Further, using Markov's inequality,
the probability that $|\lambda_2|$ is greater than
$c \lambda_{loose}(D)$, for any $c\geq 1$, is bounded by
$(1+{\rm o}(1)) c^{-\log(N^4)/\log(1/\lambda_H)}$, so that for large $N$ it is
very rare for $|\lambda_2|$ to be significantly above the loose bound
$\lambda_{loose}(D)$.

\subsection{Tight Bound}

We now tighten the bound.  On a given iteration of the Schwinger-Dyson
equations, we go from a product of $k$ traces to a product of $k+1$, $k-1$,
or $k-2$
traces.  We will keep track of how the matrices move under
this iteration process using a function $f_n((l,i))$ from pairs of integers
to pairs of integers.  We say that the matrix $U(s_{l,i})$ in the
given product of traces, $L_1 L_2 ... L_k$, is in
position $(l,i)$.  Let us consider the case of a term on the first 
line, where $m$ increases by one.  Then, for any given $j$ in the
sum on the first line, we say that
the matrix in position $(i,1)$, for $i<j$ on the $n+1$-st iteration corresponds
to the matrix in position $(1,i)$ on the $n$-th iteration, and so
$f_n((1,i))=(1,i)$, while the matrix
in position $(2,i)$ on the $n+1$-st iteration
corresponds to the matrix in position $(1,i+j-1)$ on the $n$-th iteration, so
$f_n((1,i+j-1))=(2,i)$.
The matrix in position $(l,i)$, for $2<l\leq k+1$ on the $n+1$-st iteration
corresponds to the matrix $(l-1,i)$ on the $n$-th iteration, so
$f_n(l-1,i)=(l,i)$.  We follow a
similar procedure for the other lines of Eq.~(\ref{SD2}) and if there are
cancellations, we keep track of how the matrix moves under the cancellations.

We then keep track of which matrix after $n$ iterations corresponds to
a given matrix before any iterations, by defining
$F_n((l,i))=f_n(f_{n-1}(...f_1((l,i))$ for $l=1,2$.  Let us say that the matrix
at position $(l,i)$ for $l=1,2$ is ``trivially moved" under the $n$-th
iteration of the Schwinger-Dyson equations if we are considering a
term in the equations which did not arise from $T^a  U(s_{l,i}))$;
that is, the matrix is trivially moved if it is not in position
$(1,j)$ using a term on the first or second line, or in position $(l,j)$
using a position from the third or fourth line, or in position $(1,1)$.
Let us define a ``rung cancellation of matrix $i$"
to be the case in which, for some $n$,
after the $n$-th iteration of the Schwinger-Dyson
equation we perform a series of cancellations such that the following
hold\cite{vert}.
First, a matrix in position $(l,j)$ is canceled against a matrix
in position
$(l',j')$ such that $(l,j)=F_{n-1}((1,i))$ and $(l',j')=F_{n-1}((2,m_1+1-i))$.
Second, at all previous iterations up to the $n-1$-th iteration, the
given matrix was trivially moved.  
If there is a rung cancellation of matrix $1$ on the first iteration,
then all matrices cancel and the trace is equal to unity; this is
precisely the case with $l=2,j=m$ discussed at the start of the
section ``Nontrivial Words".  Note that the matrix in position
$(l',j')=F_{n-1}((2,m_1+1-i))$ is equal to $U(s_{2,m_1+1-i})=U(s_i+D/2)$ which
is why the matrix in position $(l,j)=F_{n-1}((1,i))$ can be canceled
against this matrix.

We now make
a stronger claim: for any given $i$, the sum of all terms with a rung
cancellation of matrix $i$ is equal to unity.  To show this, consider
the trace 
${\rm tr}(U(s_{m}+D/2) ...  U(s_{i+1}+D/2) X^{\dagger} U(s_{i-1}+D/2) ... 
U(s_{1}+D/2))
{\rm tr}(U(s_{1})...U(s_{i-i}) X U(s_{i+1}) ... U(s_{m})$, 
where $X$ is some arbitrary unitary matrix.  Averaging
this trace over all unitary matrices $U(s)$ and over all unitary matrices $X$
with the Haar measure,
we find that the trace is equal to unity.  However, applying the Schwinger-Dyson
equations to this trace generates precisely the sum of terms mentioned
above, those in which there is a rung cancellation
of matrix $i$.  Thus, this sum equals unity.  We further claim that
for any given $i_1,i_2,...,i_d$, the sum of all terms with rung
cancellations of matrices $i_1,i_2,...i_d$ is equal to unity, as may
be shown by considering a trace in which matrices $U(s_{i_1}),U(s_{i_2}),...$
are replaced by $X_1,X_2,...$, and the trace is averaged over the different
$X_1,X_2,...$.

On the other hand, if a term terminates at level $n$
and matrix $i$ does not have
a rung cancellation, then at some previous iteration $n$
either the matrix was
not trivially moved or was canceled against a matrix in position
$(l,j)$ such that $(l,j)=F_{n-1}(l',j')$ with $(l',j')\neq (2,m_1+1-i)$.  
In the later case, for $l'=2$ we know that $s_{m+1-j'}+D/2=s_{i}$,
while for $l'=1$ we know that $s_{j'}=s_{i}$, thus in both
cases identifying some $k\neq i$ such
either $s_{1,i}=s_{1,k}$ or $s_{1,i}=s_{1,k}+D/2$.  If the
matrix was not trivially moved, we can also identify some
$k\neq i$ with the  same properties.  Let us write $k=\tau(i)$ in both
cases, for some function $\tau(i)$.

Now consider the sum of terms in which for no $i$ is there a rung
cancellation of matrix $i$.  By the inclusion-exclusion principle in
combinatorics, this is equal to the sum of all terms, minus the
sum over $i$ of the sum of terms in which there is a rung
cancellation of matrix $i$,
plus one half the sum over $i_1\neq i_2$ of the sum
of terms in which there are rung cancellations of
matrices $i_1,i_2$, and so on.
This is equal to the sum of all terms minus the sum
\be
m_1-m_1(m_1-1)/2+m_1(m_1-1)(m_1-2)/6-...=1.
\ee
Thus, the sum of all terms is equal to one plus
the sum of terms in which for no $i$ is there a rung
cancellation of matrix $i$.  So, we now focus on the sum of
terms with no rung cancellation, which we define to be
$E_0'(s_1,...,s_m)$.  If $s_1,...,s_m$ has a shift symmetry as above,
then there may be terms in this sum terminating at the first level;
the sum of these terms is $o-1$.

Each $E_0$ we are averaging over the unitary group results
from a particular set of choices of
$s_1,...,s_m$ in the sum in Eq.~(\ref{av}).  There are $D^m$ different
terms in this sum in (\ref{av}).  We begin by bounding, for any given level
$n$,
the number of choices of $s_1,...,s_m$ which give rise to an $E_0$
which produces a term in the Schwinger-Dyson equations which terminates
at level $n$ with no rung cancellations.  Suppose for a given
choice of $s_1,...,s_m$ there is such a term
which terminates at level $n$ with no rung cancellations.
There were two traces of $m$ unitaries in the definition of $E_0$; then,
after canceling successive pairs $U(s) U(s+D/2)$,
we have $m_1\leq m$ unitaries for some $m_1$.
The number of different initial choices of $s_1,...,s_D$ which produce
a given $m_1$ after these cancellations is given in Eq.~(\ref{nchoice}).

Then we iterate the Schwinger-Dyson equations with a
particular choice of $l,j$ at each level, where
$1\leq l \leq 2m_1$ and $1\leq j \leq 2m_1$ as given in Eq.~(\ref{SD2});
if
we pick a term from the first or second line of the Schwinger-Dyson
equations, we set $l=1$ at that level.  At each such iteration
of the Schwinger-Dyson equations, there may be cancellations in
two different traces if the term came from the second line
of Eq.~(\ref{SD2}), with at most $m_1$ cancellations in each trace, or
cancellations
in two different places of a single trace, if the term came from the fourth
line of Eq.~(\ref{SD2}), with at most $m_1$ cancellations in each place.
Let us call the number
of cancellations $c_1,c_2$ with $0\leq c_1\leq m_1$ and $0\leq c_2 \leq m_1$.
Then,
by specifying $l,j,c_1,c_2$
for each iteration, we succeed in fully specifying how the matrices
move under the $n$ iterations of the Schwinger-Dyson equation; this requires
specifying $2n$ numbers ranging from $1...2m_1$, and $2n$ numbers
ranging from $0...m_1$.
In particular,
since there are no rung cancellations, we succeed in specifying for each
$i$, $1 \leq i \leq m_1$, some $j\neq i$ such that either $s_{1,i}=s_{1,j}$
or $s_{1,i}=s_{1,j}+D/2$, giving the function $\tau(i)$.
Having specified this function,
there are now only at most $[D/(D-1)] (D-1)^{m_1/2}$
possible values of $s_{1,1},...,s_{1,m_1}$.  To show this, we start
by specifying the value of
$s_{1,1}$, which can assume any of $D$ different values.  
By specifying $s_{1,1}$ we have fixed the value of
of $s_{1,\tau(1)}$, as well as the value of any $j$ such that $\tau(j)=1$,
so that there are now at most $m_1-2$ different values of $s_{1,i}$
which remain undetermined.  We then find the smallest $j_1$ such that
$s_{1,j_1}$ is undetermined and specify its value.  Note that there are only
$D-1$ possible values of this $s_{1,j_1}$ since, by assumption, $s_{1,j_1}\neq
s_{1,j_1-1}+D/2$.  Having specified this $s_{1,j_1}$, we have fixed the value
of $s_{1,\tau(j_1)}$ as well as the value of any $j$ such that
$\tau(j)=j_1$.  We then find the smallest $j_2$ such that $s_{1,j_2}$ is
undetermined and specify that value.  Proceeding in this way, we succeed
in specifying $s_{1,1},...,s_{1,m_1}$ by specifying one of at most
$[D/(D-1)] (D-1)^{m_1/2}$ different choices.
Thus, there are at most
\be
[D/(D-1)] (D-1)^{m_1/2} (2m_1)^{2n} (m_1+1)^{2n} \leq [D/(D-1)]
(D-1)^{m_1/2} (2m_1)^{4n}
\ee
such choices of $s_1,...,s_{m_1}$ which can produce a term which terminates
at level $n$.
Using Eq.~(\ref{nchoice}),
the number of choices of $s_1,...,s_m$ which can produce a term which
terminates at level $n$ is at most
\be
\label{numnmin}
\sum_{m_1=0}^{m} 
(D-1)^{m/2} 2^m
(2m_1)^{4n}\leq 
(D-1)^{m/2} 2^m 
\frac{(2m+1)^{4n+1}}{4n+1}.
\ee

For any $s_1,...,s_m$, we define $n_{min}(s_1,...,s_m)$ to be the
smallest level at which a term terminates with no rung cancellations.
We re-write the sum in Eq.~(\ref{av}) as
\be
\label{cutn}
E_1=1+\Bigl( \frac{1}{D} \Bigr)^m \sum_{n=0}^{\infty} 
\sum_{s_1=1}^{D}
\sum_{s_2=1}^{D} ...
\sum_{s_m=1}^{D}
\delta_{n_{min}(s_1,...,s_m),n} E_0'(s_1,...,s_m),
\ee
so that the second sum is over the set of all values of $s_1,...,s_m$ with
the given $n_{min}=n$.  We note that the bound of Eq.~(\ref{valBd}) continues
to apply to the terms terminating with no rung cancellations, and
the bound of Eq.~(\ref{numBd}) continues to bound the number of such terms
terminating with no rung cancellations.  From Eq.~(\ref{valBd}),
a bound on the value of the term
terminating at the $n$-th level, for any $n\geq 0$ is
\be
N^2 N^{-(2/3) n}.
\ee
Therefore, for any
$s_1,...,s_m$,
\begin{eqnarray}
\label{e0nmin}
E_0'(s_1,...,s_m)
&\leq& N^2 \sum_{n\geq n_{min}(s_1,...,s_m)} N^{-(2/3) n}
(2m-1)^n
\\ \nonumber
&=& N^2 \frac{[N^{-2/3} (2m-1)]^{n_{min}}}{1-N^{-2/3} (2m-1)}.
\end{eqnarray}
From Eqs.~(\ref{numnmin},\ref{cutn},\ref{e0nmin}),
\begin{eqnarray}
E_1& \leq & 1+N^2 \lambda_H^m \sum_{n=0}^{\infty} 
\frac{(2m+1)^{4n+1}}{4n+1}
\frac{[N^{-2/3} (2m-1)]^n}{1-N^{-2/3} (2m-1)}
\\ \nonumber
&\leq &
1+N^2 \lambda_H^m 
\sum_{n=0}^{\infty} \frac{2m+1}{(4n+1)[1-N^{-2/3}(2m-1)]}
[N^{-2/3} (2m+1)^5]^n.
\end{eqnarray}

We then pick $m=(1/4) N^{2/15}$, so that
$N^{-2/3} (2m+1)^5\leq 1/2$ and
\begin{eqnarray}
|\lambda_2|& \leq &(E_1-1)^{1/m} \leq N^{2/m} \lambda_H (1+{\rm O}(1))^{1/m}
\\ \nonumber
&=&\lambda_H (1+{\rm O}(\log(N) N^{-2/15}).
\end{eqnarray}
As before, using Markov's inequality,
the probability that $|\lambda_2|$ is greater than
$c\lambda_{H}(D)$, for any $c\geq 1$, is bounded by
$c^{-(1/4) N^{2/15}}
(1+{\rm O}(\log(N) N^{-2/15})$.

This shows that for any $\epsilon$, the probability that
$\lambda_2\leq \lambda_H+\epsilon$ approaches unity as $N\rightarrow\infty$.
Combined with the previous lower bound (\ref{lowB}), this proves the main
result.

\section{Discussion}
We consider some analogies between these results and lattice gauge
theory, some applications of these results, and some extensions.
We begin with
analogies between
the random construction of quantum expanders and
lattice gauge theory and the Eguchi-Kawai construction\cite{ek}.

\subsection{Gauge Theory Analogies}
Consider a lattice gauge $U(N)$
theory in $D/2$ dimensions on a hypercubic lattice,
with unitary matrices $U_{\hat d}(x)$ defined for each link of the lattice.
Here, $x$ represent a point on the lattice, and $\hat d$ represents the
direction of the link:
if $d\leq D/2$, it points in the direction of increasing the $d$-th
coordinate by unity, while if $D/2<d\leq D$, then it points in the
direction of decreasing
$d-D/2$-th coordinate by unity.  
Then, for a given
choice of $s_1,...,s_m$ we can define a path, starting at the origin, and
then moving in direction $\hat s_1,\hat s_2,...$ until $m$ steps have been
taken.  We can define a product of
traces associated with this path: ${\rm tr}(U_{s_1}(0) U_{s_2}(0+\hat s_1) ...)
{\rm tr}(... U_{s_2}(0+\hat s_1)^{\dagger} U_{s_1}(0)^{\dagger})$.
For certain choices of the $s_1,...$ this path returns to the origin
after $m$ steps, in which case the product of traces is a product
of two Wilson loops.  If, however, the path does not return to the
origin, the product of traces is not invariant under non-Abelian
gauge transformations, and hence the average of the product of traces
is equal to unity.

At infinite coupling, all of the unitary matrices are independent,
except for the constraint $U_{\hat d}(x)=U_{\hat{d+D/2}}(x)$, and even
if the path does return to the origin,
the average of this product of traces is equal to unity, unless, by
chance, the path of length $m$ exactly retraces itself.
The probability
of this retracing, for a random path, is precisely the Cayley tree return
probability discussed previously.  Thus, this lattice gauge theory
at infinite coupling has
${\rm tr}(U_{s_1}(0) U_{s_2}(0+\hat s_1) ...)
{\rm tr}(... U_{s_2}(0+\hat s_1)^{\dagger} U_{s_1}(0)^{\dagger})=
1+N^2 D^{-m} N(0,m)$.
The Eguchi-Kawai construction is an approximation to large
$N$ gauge theory which replaces the infinite lattice by a single site:
this turns
${\rm tr}(U_{s_1}(0) U_{s_2}(0+\hat s_1) ...)
{\rm tr}(... U_{s_2}(0+\hat s_1)^{\dagger} U_{s_1}(0)^{\dagger})$ into
$E_0(s_1,s_2,...)$, the quantity we considered before.  Thus, this paper
can be seen as an estimation of corrections to the Eguchi-Kawai
construction in the infinite coupling limit.  There are a number interesting
terms in these corrections: for example, the average ${\rm tr}(U(1) U(1))
{\rm tr}(U^{\dagger}(1) U^{\dagger}(1))$ is equal to 2 as calculated before,
but the corresponding average in the lattice gauge theory is equal to 1.

\subsection{Applications}
The general properties of ground states of local Hamiltonians with an
excitation gap have become of great interest recently.  A basic
result\cite{loc1,loc2} is that correlations decay exponentially in
such systems.
One application of quantum expanders is to finding matrix product
states of one-dimensional quantum systems with the following
properties: the correlation length is of order unity, the Hilbert space
dimension on a single site is small, also of order unity, and yet the
entanglement entropy across any cut is large.  
As an example, consider a matrix product state of the form:
\be
\label{mps}
\Psi(s_1,s_2,...s_N)=\sum_{\alpha,\beta,...}
A_{\alpha,\beta}(s_1)
A_{\beta,\gamma}(s_2)
A_{\gamma,\delta}(s_3)...
\ee
where $s_1,s_2,...,s_N$ are spin variables in a one-dimensional quantum
system of $N$ sites.
Associated with the matrix product state is a completely positive map as
in Eq.~(\ref{mapdefn}).  If this map has a gap in its eigenspectrum to the
second largest eigenvalue, then the state $\Psi$ has exponentially
decaying correlations\cite{mps}, so that if operator $A$ has support
on sites $1,...,j$ and operator $B$ has support on sites $j+l,...,N$, then
$\langle \Psi,A B \Psi \rangle-\langle \Psi,A \Psi \rangle\langle \Psi, B
\Psi \rangle$ is exponentially small in $l$, as required for the ground
state of a gapped, local quantum system.
However, as discussed in \cite{em1}, this means that the
existence of quantum expanders implies
that the mere fact of exponentially decaying correlations does not
suffice to prove bounds on entanglement entropy.  Instead,
bounds on the entanglement entropy\cite{areal} proceed through a different
route and currently give weak bounds.

However, in \cite{areal}, a conjecture was developed regarding properties
of quantum expanders that may help in proving tighter bounds
on entanglement entropy.  Consider the following different correlation
function.
Let $A$ be an operator with support on the sites $1,2...,j-l$ and
$j+l,j+l+1,...,N$.
Let
$\Psi=\sum_{\alpha=1} A(\alpha) \Psi_{L}(\alpha)\otimes
\Psi_{R}(\alpha)$, where $\Psi_{L}(\alpha)$ are orthonormal
states on sites $1,...,j$ and $\Psi_{R}(\alpha)$ are orthonormal states
on $j+1,...,N$.
Let
$B_L=\sum_{\alpha=1} O(\alpha) \Psi_{L,0}(\alpha) \rangle\langle \Psi_{L,0}
\otimes \openone_R$, where $\openone_R$ is the unit operator on $X_{j+1,N}$,
for some function $O(\alpha)$.
Then, it was shown that for a gapped local Hamiltonian,
\be
\label{fwdback}
\langle \Psi_0,A B_L \Psi_0 \rangle-\langle \Psi_0,A \Psi_0\rangle
\langle\Psi_0, B_L\Psi_0\rangle \leq
\Vert A \Vert \Vert B \Vert {\cal O}(\exp[-l/l_0]),
\ee
for some $l_0$.

It was conjectured in \cite{areal} that there is a function $f(D_{eff})$
such that if Eq.~(\ref{fwdback}) holds for
a state $\Psi$ for some $l_0$,
then the entanglement entropy of $\Psi$ is bounded by
$f(D^{l_0})$.  
Interestingly, it seems that an expander where the $A(s)$
are random unitaries is unlikely to satisfy Eq.~(\ref{fwdback}).
If this could be shown to be a general property of expanders, showing
the conjecture, this would provide another way of studying area laws in quantum
systems.

\subsection{Extensions}
The method of Schwinger-Dyson equations used here is fairly general
and could be applied to other groups, such as $O(N)$ or $Sp(2N)$.  We have
not done the calculation, but it seems that random choices from these groups
will also give quantum expanders.  Always, the unit matrix is an eigenvector
of the map ${\cal E}(M)$ with eigenvalue unity.  Any matrix in the center
of the group is also an eigenvector of ${\cal E}(M)$ with eigenvalue
unity, but for these cases, all elements of the center are proportional to
the identity matrix, and thus do not give rise to additional eigenvectors with
unit eigenvalue.

The method can be directly extended to the non-Hermitian case.  Some of the
combinatorics become slightly easier here.  From
Eq.~(\ref{tr}), the average of $\sum_{a=1}^{N^2} |\lambda_a|^{2m}$
over the unitary group is bounded by the average of the trace:
\begin{eqnarray}
\label{trnHav}
&\Bigl(\frac{1}{D}\Bigr)^{2m} \sum_{s_1=1}^D
...
\sum_{s_{m}=1}^D
\sum_{\overline s_1=1}^D
...
\sum_{\overline s_m=1}^D
E[&
{\rm tr}(U(\overline s_1) U(\overline s_2) ... U(\overline s_m)
U^{\dagger}(s_m) ...
U^{\dagger}(s_2) U^{\dagger}(s_1)) \\ \nonumber
&& \times
{\rm tr}(U(s_1) U(s_2) ... U(s_m)U^{\dagger}(\overline s_m)
... U^{\dagger}(\overline s_2) U^{\dagger}(\overline s_1))]
\end{eqnarray}
The probability of having 
$U(s_1) U(s_2) ... U(s_m)U^{\dagger}(\overline s_m)
... U^{\dagger}(\overline s_2) U^{\dagger}(\overline s_1)$ cancel to
the identity matrix
is equal to $1/D^{m}$.  
Let
\be
\lambda_{nH}=\frac{1}{\sqrt{D}}.
\ee
Carrying through the calculation one
finds that, for any $\epsilon>0$, the probability that 
$|\lambda_2|\leq \lambda_{nH}+\epsilon$
approaches unity as
$N\rightarrow \infty$.  Note that $\lambda_{nH}< \lambda_{H}$ and
also note that in the
non-Hermitian case even a tight estimate on the average
of the trace only provides an upper bound on the eigenvalue, due to
the inequality in Eq.~(\ref{tr}).  However, numerical work suggests that
the eigenvalue is asymptotically equal to $\lambda_{nH}$ with high probability
in this case.

We can also provide a lower bound on the trace in the non-Hermitian case.
For any choice of unitaries $U(s)$,
the sum of terms in (\ref{trnHav}) with
$s_i=\overline s_i$ is bounded below by $1/D^m$, while the other terms
are all positive, so that the given average (\ref{trnHav}) is bounded
below by $N^2 D^{-m}$.  This result extends readily to an arbitrary
choice of $A(s)$, constrained only by the trace-preserving condition,
$\sum_{s=1}^D A(s) A^{\dagger}(s)=\openone$.

{\it Acknowledgments---}  I thank T. J. Osborne
for many useful discussions
on this topic.  This work was inspired by discussions with
A. Harrow, T. J. Osborne, and F. Verstraete at the
workshop on ``Lieb-Robinson Bounds and Applications" at the Erwin
Schr\"{o}dinger Institute.
This work supported by U. S. DOE Contract No. DE-AC52-06NA25396.

\appendix
\section{Quantum Edge Expanders}

In the appendix, we discuss the relationship between quantum expanders
and another concept, a quantum version of an edge expander.
We define a
map to be a ``quantum edge expander" if the following condition holds:
for any $N$-by-$N$ Hermitian matrix $P$ such that $P^2=P$ and such that
$P$ has $l$ non-zero eigenvalues, $l\leq N/2$,
\be
\label{qee}
{\rm tr}(P {\cal E}(P))\leq \lambda_e {\rm tr}(P),
\ee
for some $\lambda_e$ less than one.  We then prove
a relation between $\lambda_e$ and $|\lambda_2|$, showing that
a quantum edge expander is a quantum expander.  This is a quantum
analogue
of a theorem of Tanner\cite{t} and Alon and Milman\cite{am}, which shows that an
edge expander has a spectral gap.  We consider only the Hermitian
case in this appendix, leaving the behavior in the non-Hermitian case
open.  We also assume that the second largest eigenvalue
is positive; the case where it is negative can be considered by looking
at the square of the map ${\cal E}(M)$.

Let $X$ be the eigenvector of ${\cal E}$ with eigenvalue $\lambda_2$.
We work in a basis in which $X$ is diagonal,
\be
X=\begin{pmatrix}
e_1 && \\
&e_2 & \\
&&...\\
\end{pmatrix}
\ee
and such that $e_1\geq e_2 \geq ... \geq e_N$.
Since $X$ is orthogonal to the unit matrix, using the inner
product $(X,N)={\rm tr}(XN)$, we have ${\rm tr}(X)=0$.
Define $m$ such that $e_i>0$ for $i\leq m$ and $e_i\leq 0$
for $i>m$.  Without loss of generality we may suppose that $m\leq N/2$,
as otherwise we could have considered the matrix $-X$ which has the same
eigenvalue.
Define $M(i,j)$ to be the matrix with an a unit entry in the $i$-th row
and $j$-th column and zero everywhere else.  Define
\be
P_{ij}={\rm tr}(M(i,i) {\cal E}(M(j,j))).
\ee
Then,
since the map ${\cal E}$ is trace preserving, we have
\be
\label{tprsrv}
\sum_i P_{ij}=1
\ee
for all $j$.  Also, we have $P_{ij}=P_{ji}$.
Finally, since ${\cal E}$ is completely positive, we have
$P_{ij}\geq 0$ for all $i,j$.
Then,
\begin{eqnarray}
\lambda_2={\rm tr}(X{\cal E}(X))
=\sum_{i=1}^N\sum_{j=1}^N e_i e_j P_{ij}.
\end{eqnarray}
Define $f_i$ by
\begin{eqnarray}
f_i=e_i/\sqrt{\sum_{i=1}^m e_i^2}\quad i\leq m\\ \nonumber
f_i=0\quad i>m.
\end{eqnarray}
Then,
\be
\label{varnl}
\lambda_2\leq \sum_{i=1}^m \sum_{j=1}^m f_i f_j P_{ij}.
\ee

Then,
\begin{eqnarray}
\label{step1}
\sum_{i=1}^m \sum_{j>i}^m (f_i^2-f_j^2) P_{ij}&=&
\sum_{i=1}^m \sum_{j>i}^m [\sqrt{P_{ij}}(f_i-f_j)]
[\sqrt{P_{ij}}(f_i+f_j)]\\ \nonumber&\leq&
\sqrt{\sum_{i=1}^m \sum_{j>i}^m P_{ij}(f_i-f_j)^2}
\sqrt{\sum_{i=1}^m \sum_{j>i}^m P_{ij}(f_i+f_j)^2}
\\ \nonumber
&\leq&
\sqrt{\sum_{i=1}^m \sum_{j>i}^m P_{ij}(f_i-f_j)^2}
\sqrt{(1/2)\sum_{i=1}^m \sum_{j=1}^m P_{ij} 2(f_i^2+f_j^2)}
\\ \nonumber
&=&\sqrt{2} \sqrt{\sum_{i=1}^m \sum_{j>i}^m P_{ij}(f_i-f_j)^2}
\leq \sqrt{2} \sqrt{(1-\lambda_2)}.
\end{eqnarray}
where the first inequality uses Cauchy-Schwarz, the last
equality uses Eq.~(\ref{tprsrv}), and the last inequality
uses Eq.~(\ref{varnl}).

Let $P_i$ be the projector onto the
vector space spanned by the first $i$ eigenvectors
of $M$.  Then,
\be
\label{step2}
\sum_{i=1}^m \sum_{j>i}^m (f_i^2-f_j^2) P_{ij}=
\sum_{i=2}^m (f_i^2-f_{i-1}^2) {\rm tr}((\openone-P_i){\cal E}(P_i)).
\ee
Using the property of a quantum edge expander, (\ref{qee}),
we have
\begin{eqnarray}
\label{step3}
\sum_{i=1}^m (f_i^2-f_{i+1}^2) {\rm tr}((\openone-P_i){\cal E}(P_i))
&\geq & \sum_{i=1}^m (f_i^2-f_{i+1}^2) {\rm tr}(P_i) (1-\lambda_e)
=
\sum_{i=1}^m (f_i^2-f_{i+1}^2) i (1-\lambda_e)
\\ \nonumber
& = &
\sum_{i=1}^m f_i^2 (1-\lambda_e)
=(1-\lambda_e).
\end{eqnarray}
Combining Eqs.~(\ref{step1},\ref{step2},\ref{step3}), we find that
\be
1-\lambda_e \leq \sqrt{2(1-\lambda_2)}.
\ee

We finally show the converse result, that a quantum expander is a quantum edge
expander.  The normalized eigenvector with unit eigenvalue is
$v_1\equiv (1/\sqrt{N}) \openone$.  We have $(v_1,P)={\rm tr}(P)/\sqrt{N}$.
So, $P={\rm tr}(P)\openone /N+P'$, where
$P'=P-{\rm tr}(P)\openone/N$.
Then ${\rm tr}(P {\cal E}(P)) \leq |\lambda_2| {\rm tr}(P'^2)
+(v_1,P)^2
=|\lambda_2| ({\rm tr}(P)-{\rm tr}(P)^2/N)
+{\rm tr}(P)^2/N$.  If ${\rm tr}(P)=l\leq N/2$, then
${\rm tr}(P)^2\leq (N/2) {\rm tr}(P)$ and
${\rm tr}(P {\cal E}(P)) \leq
|\lambda_2| {\rm tr}(P)+(1-|\lambda_2|){\rm tr}(P)/2$ so
\be
\lambda_e \leq |\lambda_2|/2+1/2.
\ee


\begin{thebibliography}{99}
\bibitem{formal} An appropriate formal definition may be to define
a $(D,|\lambda_2|,N,S)$ expander, where $S$ is the entropy of the
eigenvector with eigenvalue unity.
\bibitem{example} An example where the series has an infinite number
of non-vanishing coefficients is the average
of ${\rm tr}(U U V V)
{\rm tr}(V^{\dagger} V^{\dagger} U^{\dagger} U^{\dagger})$ over all
unitary matrices $U,V$.  A single application of Eq.~(\ref{SD2}) gives
$1+(1/N) E[{\rm tr}(U V V U^{\dagger} V^{\dagger} V^{\dagger})]
-(1/N) E[{\rm tr}(U) {\rm tr}(U V V) {\rm tr}(V^{\dagger} V^{\dagger}
U^{\dagger} U^{\dagger})]$.
Applying Eq.~(\ref{SD2}) again gives 
$1+(1/N^2) E[{\rm tr}(V V) {\rm tr}(V^{\dagger} V^{\dagger})]
+(1/N^2)
E[{\rm tr}(U U V V)
{\rm tr}(V^{\dagger} V^{\dagger} U^{\dagger} U^{\dagger})]-(2/N^2)
E[{\rm tr}(U V V){\rm tr}(V^{\dagger} V^{\dagger} U^{\dagger}]$.
This is equal to $1+(2/N^2) 
+(1/N^2)
E[{\rm tr}(U U V V)
{\rm tr}(V^{\dagger} V^{\dagger} U^{\dagger} U^{\dagger})]-(2/N^2)$.
Thus, 
$E[{\rm tr}(U U V V)
{\rm tr}(V^{\dagger} V^{\dagger} U^{\dagger} U^{\dagger})]=1+(1/N^2)
E[{\rm tr}(U U V V)
{\rm tr}(V^{\dagger} V^{\dagger} U^{\dagger} U^{\dagger})]=1+(1/N^2)+(1/N^4)+...$.
\bibitem{vert} We refer to these as ``rung cancellations" because
in certain diagrammatic expansions they correspond to a single
rung of a ladder diagram
from field theory.
\end{thebibliography}
\end{document}